\documentclass[iop]{emulateapj}  

\usepackage{natbib}
\usepackage{amssymb}
\usepackage{amsmath}
\usepackage{amstext}
\usepackage{xcolor}
\usepackage[varg]{txfonts}	
\usepackage{microtype}		
\usepackage{graphicx}
\usepackage{multirow}

\let\url\relax
\makeatletter
\renewcommand{\fps@figure}{tp}
\makeatother
\setkeys{Gin}{width=\linewidth,height=0.9\textheight,keepaspectratio=true}

\newcommand\Jup{\ensuremath{_{\mathrm{\scriptscriptstyle Jup}}}} 
\newcommand\disk{\ensuremath{_{\mathrm{\scriptscriptstyle disk}}}} 
\newcommand\EUV{\ensuremath{_{\mathrm{\scriptscriptstyle EUV}}}}
\newcommand\FUV{\ensuremath{_{\mathrm{\scriptscriptstyle FUV}}}} 
\newcommand\EGG{\ensuremath{_{\mathrm{\scriptscriptstyle EGG}}}}

\slugcomment{The Astrophysical Journal, XXX:XXXX--XXXX, 2016 XX XX}

\shorttitle{Disks in the Carina nebula}
\shortauthors{Mesa-Delgado et al.}

\begin{document}


\title{Protoplanetary disks in the hostile environment of Carina}


\author{A.~Mesa-Delgado\altaffilmark{1$\dagger$}, L.~Zapata\altaffilmark{2}, W.J.~Henney\altaffilmark{2}, T.H.~Puzia\altaffilmark{1}, \& Y.G.~Tsamis\altaffilmark{3}}
\affil{\altaffilmark{1}Instituto de Astrof\'isica, Facultad de F\'isica, Pontificia Universidad Cat\'olica de Chile, Av.~Vicu\~na Mackenna 4860, 782-0436 Macul, Santiago, Chile}
\affil{\altaffilmark{2}Instituto de Radioastronom\'ia y Astrof\'isica, Universidad Nacional Aut\'onoma de M\'exico, Campus Morelia, Apartado Postal 3-72, 58090 Morelia, Michoac\'an, M\'exico}
\affil{\altaffilmark{3}Department of Physics and Astronomy, University College London, London WC1E~6BT, United Kingdom}
\email{$\dagger$E-mail: amesad@astro.puc.cl}


\begin{abstract}
We report the first direct imaging of protoplanetary disks in the star-forming region of Carina, the most distant, massive cluster in which disks have been imaged. 
Using the Atacama Large Millimeter/sub-millimeter Array (ALMA), disks are observed around two young stellar objects (YSOs) that are embedded inside evaporating gaseous globules and exhibit jet activity. 
The disks have an average size of 120~AU and total masses of 30 and 50~\(M\Jup\).
Given the measured masses, the minimum timescale required for planet formation (\(\sim1-2\)~Myr) and the average age of the Carina population (\(\sim1-4\)~Myr), it is plausible that young planets are present or their formation is currently ongoing in these disks.
The non-detection of millimeter emission above the 4$\sigma$ threshold (\(\sim 7 M\Jup\)) in the core of the massive cluster Trumpler~14, an area containing previously identified proplyd candidates, suggest evidence for rapid photo-evaporative disk destruction in the cluster's harsh radiation field.  
This would prevent the formation of giant gas planets in disks located in the cores of Carina's dense sub-clusters, whereas the majority of YSO disks in the wider Carina region remain unaffected by external photo-evaporation.
\end{abstract}

\section{Introduction} \label{intro}  
It is generally accepted that most stars are born in dense clusters \citep{blaauw64, ladalada03}, where massive and low-mass stars co-exist and evolve. 
In these regions, the far ultraviolet (FUV) and extreme ultraviolet (EUV) radiation from the most massive stars can curtail disk lifetimes of the low-mass protostars via external photo-evaporation, caused by the heating and dissociating effects of ionizing radiation \citep[e.g.][]{johnstoneetal98, adamsetal04, clarke07, gortietal15}. 
Depending on the intensity and the duration of the exposure, photo-evaporation by external irradiation can potentially lead to disk dispersal on timescales shorter than the minimum \(1-2\)~Myr required for planet formation \citep{hubickyjetal05, lissaueretal09, najitakenyon14, almapartneretal15}.
Although internal photo-evaporation due to FUV and x-ray radiation from the protostar itself may be important, this effect generally acts on longer timescales of about $3-5$~Myr \citep[e.g.][]{owenetal12, gortietal15}. 
Understanding the impact of external photo-evaporation on disks within massive clusters is key to assessing their chances for survival and their planet-formation capacity.
   
\cite{richertetal15} found no evidence for significant disk dispersal by analyzing YSOs in a large sample of stellar associations.
Due to stellar crowding and/or PAH contamination, their analysis is inconclusive for the cores of the massive clusters in Orion, M~17, and Carina, exactly where external photo-evaporation could be important.
A clear example is Orion's rich population of proplyds \citep{odelletal93}, where millimeter observations show that disks have been eroded down to \(\sim 3 M\Jup\) within 0.03~pc from the Trapezium cluster \citep[e.g.][]{eisneretal08, mannwilliams10, mannetal14}.

Whereas the Orion nebula is dominated by a single O7V-type star \citep[\(\theta^1\)\,Ori~C;][]{simondiazetal06}, the Carina nebula, at
a distance of \(\sim 2.3\)~kpc  \citep[][hereinafter SB08]{smithbrooks08}, hosts nearly one hundred O-type stars and tens of thousands of lower-mass young stars \citep{povichetal11, povichetal11b, preibischetal11, preibischetal11b}. 
The clusters Trumpler (Tr) 14 and 16 have ultraviolet (UV) luminosities about 20 and 60 times higher than \(\theta^1\)\,Ori~C, and they are home of the most massive and luminous stars in the Galaxy (see review by SB08). 
As a prototype of massive star-forming regions in our Galaxy, Carina is ideally suited to an investigation of disk evaporation: its older age (\(1-4\)~Myr; SB08) could allow enough time for a deeper action of the external photo-evaporation process, especially in the core of the massive clusters.

In this Letter, we present novel observations with the Atacama Large Millimeter/sub-millimeter Array (ALMA) of YSO disks in Carina and investigate the implications for disk survival in the face of external photo-evaporation.

\section{Observations and data reduction} \label{obsred}  
\begin{figure*}
   \centering
   \includegraphics[width=\linewidth]{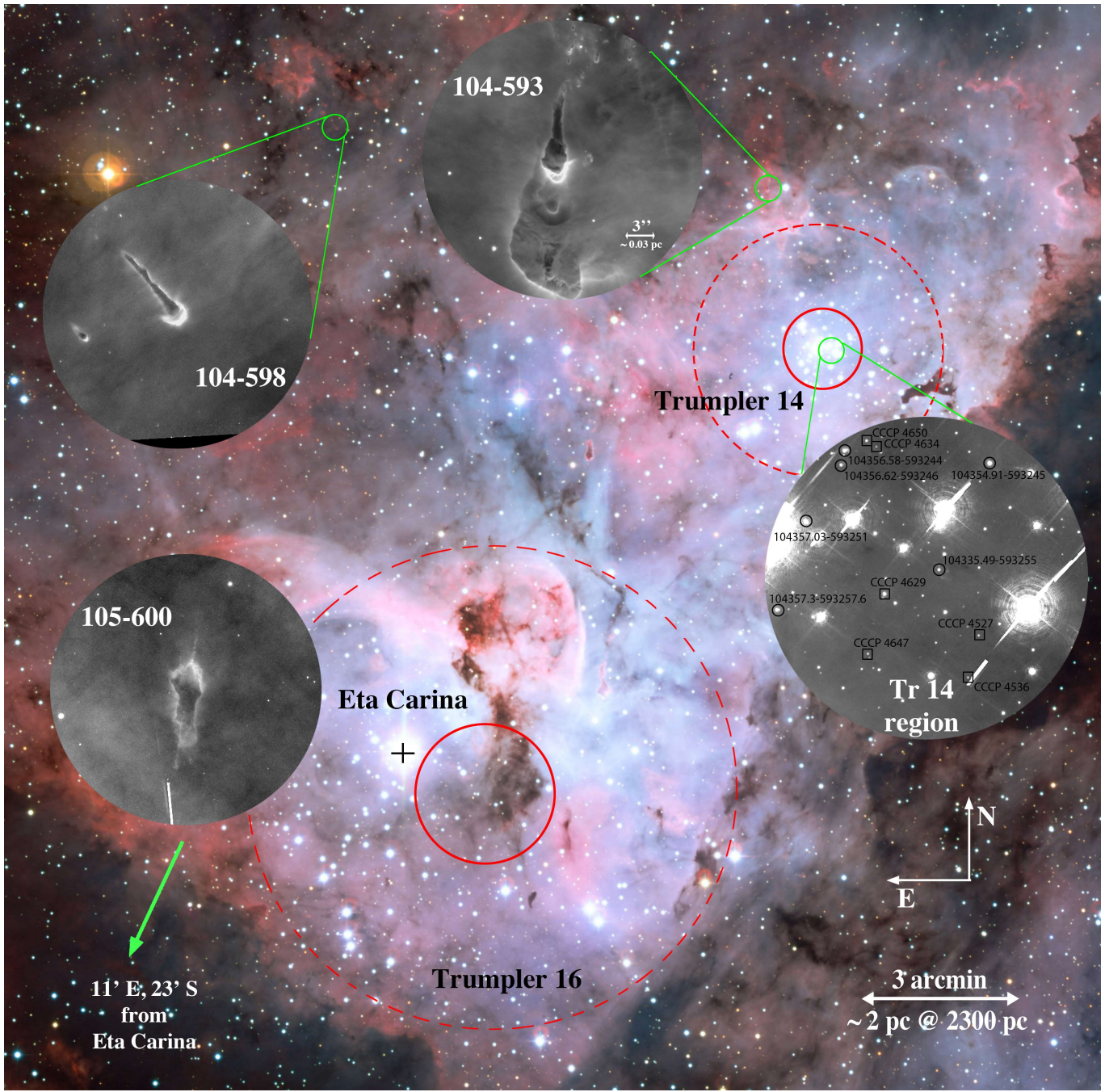} 
   \caption{
The ALMA fields and the clusters Tr~16 and Tr~14 are shown on a public composition of the Carina nebula. 
Red circles mark upper/lower limits to the size of irradiation zones around each cluster (see \S\ref{photo}). 
The ALMA fields are represented by the close-up images observed with the \textit{Hubble Space Telescope} and its Advanced Camera for Surveys instrument in the F658N narrow-band filter. 
Within the Tr~14 region, the circles contain proplyd candidates, whereas the squares mark sources with near-infrared excess. 
}
   \label{f1}
 \end{figure*}

We observed four regions covering a range of separations from Tr~16 and Tr~14, centered on sources identified by \cite{smithetal03} as potential protoplanetary disks (see Fig.~\ref{f1}).
We focused on the evaporating gas globules (EGGs) named as 104-593, 105-600 \citep[frEGG-1;][]{sahaietal12}, and 104-598, as well as a section of the Tr~14 core that contains the best candidates for bona fide proplyds in Carina \citep{smithetal10}. 

\begin{deluxetable*}{cccccccccc}
\tabletypesize{\small}
\tablecaption{\label{flux}
Physical parameters of the protoplanetary disks detected in Carina.}
\tablecolumns{10}
\tablewidth{0pc}
\tablehead{
\colhead{}& \multicolumn{2}{c}{Position$^a$} & \colhead{Flux Density$^b$} & \colhead{Spectral$^c$} &\multicolumn{3}{c}{De-convolved Size$^d$} & \multicolumn{2}{c}{Disk Mass$^e$}\\
\colhead{}& \colhead{$\alpha_{2000}$} & \colhead{$\delta_{2000}$} & \colhead{229~GHz} & \colhead{Index} & \colhead{Major} &  \colhead{Minor} & \colhead{P.A.} & \colhead{40 K}& \colhead{20 K}\\
\colhead{Source} &  \colhead{($^h$ $^m$ $^s$)} & \colhead{($^\circ$ $'$ $''$)} & \colhead{(mJy)} & \colhead{}  & \multicolumn{2}{c}{(milli-arcsecond)} & ( $^\circ$) & \multicolumn{2}{c}{(\(M\Jup)\)}
}
\startdata
  104-593 & 10 44 05.360 & $-$59 29 40.77 & $1.5\pm0.2$ & 2.9 & $45\pm7$ & $25\pm5 $ & $17\pm4$ & $45$ & $105$\\
  105-600 & 10 46 32.905 & $-$60 03 53.67 & $0.90\pm0.09$ & 2.5 & $50\pm5$ & $30\pm6$ & $110\pm12$ & $27$ & 62 \\
  104-598 & 10 45 14.200 & $-$59 28 25.00 & $\leq 0.24^f$ & -- & -- & -- & -- & $\leq 7$ & $\leq 16$\\
  Tr~14 region   &  10 43 55.490  & $-$59 32 55.00 &  $\leq 0.24^f$ & -- & -- & -- & -- & $\leq 7$ & $\leq 16$
\enddata
\tablenotetext{a}{Positional errors are 0.0075$^s$ for RA and 0.005$''$ for DEC.}
\tablenotetext{b}{Obtained from averaging all the four spectral windows.}
\tablenotetext{c}{From the density fluxes between the 220 and 237 GHz spectral windows.}
\tablenotetext{d}{From the \textsc{gaussfit} task of the CASA package.}
\tablenotetext{e}{See \S\ref{results}.}
\tablenotetext{f}{4$\sigma$ upper limit.}
\label{tab1}
\end{deluxetable*}

The observations were performed on 2015 October 27 as part of the Cycle~3 program 2015.1.01323.S.
We used 45 of the 12-m antennas, yielding baselines with projected lengths from 21 to \(16,196\)~m. 
The positions of the phase centers of each pointing are given in Table~\ref{tab1}.   
Continuum images were obtained in Band~6, with integration times of 160 s and a frequency coverage of \(220-238\)~GHz within a primary beam with a full width half maximum of about \(27''\).
The digital correlator was configured to four wide spectral windows in dual polarization mode centered at \(223.99\) GHz,  \(234.07\) GHz, \(220.87\) GHz, and  \(237.02\) GHz (average frequency of \(228.5\)~GHz, corresponding to \(\lambda \simeq 1.3\)~mm). 
The effective bandwidth for the 223 and 234~GHz spectral windows was 2000~MHz, and 938~MHz for the 220 and 237~GHz windows.
The weather conditions were very good and stable, with an average precipitable water vapor of 1.06~mm and system temperature of 60 to 200~K.
The ALMA calibration included simultaneous observations of the 183~GHz water line with water vapor radiometers, used to reduce the atmospheric phase noise.
Quasars J1107\(-\)4449 and J1049\(-\)6003 were used to calibrate the bandpass and the gain fluctuations, respectively. 

Data reduction was performed using the Common Astronomy Software Applications package \citep[CASA 4.2.2; ][]{mullinetal07}. 
Imaging of the calibrated visibilities was done using the task \textsc{clean}. 
The resulting root-mean-square noise for the continuum emission ($\sigma$) was about 6~$\mu$Jy at an angular resolution of $0.03'' \times 0.02''$ with a position angle (P.A.) of $-46.34^\circ$.
We thus resolve structures with physical sizes larger than 70~AU~\(\times\)~45~AU, or roughly the size of the Solar System.
In deriving the integrated fluxes, we used a Briggs robustness parameter of 0.5 in the {\sc  clean} task to obtain a better signal-to-noise ratio at the cost of losing some spatial resolution. 
We averaged all line-free channels in the four spectral windows for the construction of the 1.3~mm continuum images.

\section{Disks within Carina's EGGs} \label{results}  
In 104-593 and 105-600 we detect disks embedded within the core of the optically opaque EGGs (Fig.~2).
Table~\ref{tab1} gives the sizes and integrated fluxes of the disks determined with the \textsc{gaussfit} task of CASA from a natural-weighted image and within a \(0.2"\times0.2"\) box centered on the detection.
The disks have peak signals about \(50\sigma\) and \(100\sigma\), and both are marginally resolved with an average de-convolved size of \(0.05''\times0.03'' \approx 120\)~AU~\(\times\)~70~AU. 
Assuming a geometrically thin, circular disk, a radius \(R\disk\) of \(\sim 60\)~AU is implied in both cases, inclined at \(\sim 35^\circ\) to the line of sight. 

\begin{figure*}
   \centering
   \includegraphics[width=\linewidth]{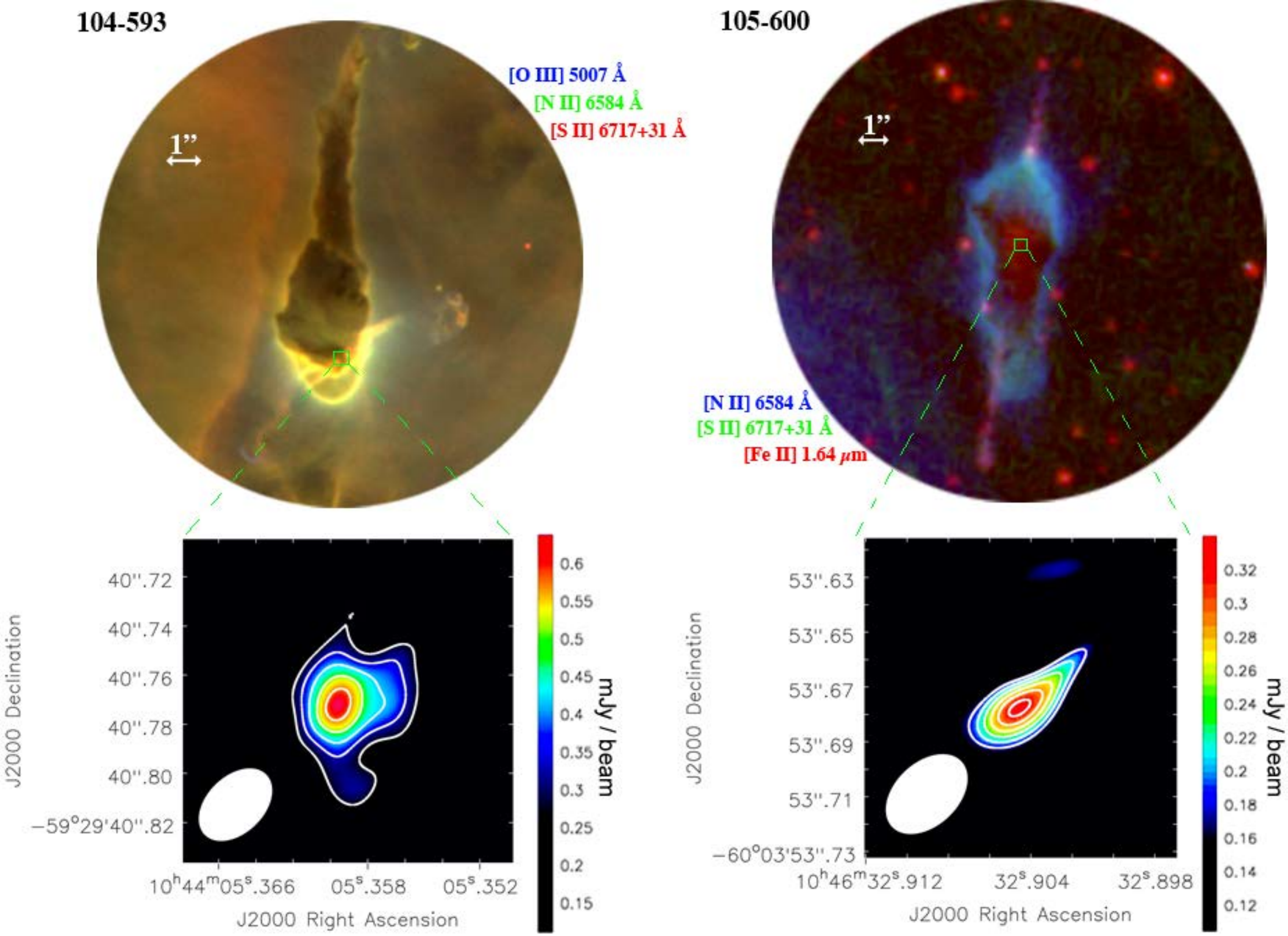} 
   \caption{
Top: RGB images of the EGGs 104-593 (left-hand side) and 105-600
(right-hand side) constructed from narrow-band \textit{Hubble Space Telescope} images
from Wide Field Canera~3 (WFC3), Advanced Camera for Surveys (ACS),
and Wide Field/Planetary Camera~2 (WFPC2).
For 104-593, WFC3 visual-band filters in lines of [O~{\sc iii}] (blue), [N~{\sc ii}] (green) and [S~{\sc ii}] (red).
For 105-600, ACS visual-band [N~{\sc ii}] filter (blue), WFPC2 visual-band [S~{\sc ii}] filter (green), and WFC3 infrared [Fe~{\sc ii}] filter (red).      
The images also show the Herbig-Haro objects HH1066 in 104-593 and HH1006 in 105-600.
Bottom: ALMA detections at $1.3$~mm continuum emission.
White contours represent values at \(38\sigma\), \(53\sigma\), \(68\sigma\), \(83\sigma\) and \(98\sigma\) mJy~beam$^{-1}$ for 104-593, and \(29\sigma\), \(34\sigma\), \(39\sigma\), \(44\sigma\), \(50\sigma\) and \(55\sigma\) mJy~beam$^{-1}$ for 105-600.
The white ellipses show the beam size of \(0.03\arcsec \times0.02\arcsec\).
}
   \label{f2}
 \end{figure*}

The disk nature of these ALMA detections is further supported by association with mid-infrared sources \citep{povichetal11}. 
Our positions match the YSOs PCYC~429 for 104-593 and PCYC~1173 for 105-600 within the errors; these are Herbig AeBe stars with masses of \(2.8\pm1.6\)~\(M_\odot\) and \(1.8\pm0.9\)~\(M_\odot\), respectively.
The mid-infrared spectra of both targets imply that they are Class 0/I sources, indicating that they still retain a massive envelope \citep{povichetal11, reitersmith13}. 
Evidence for ongoing accretion onto the protostars is provided by the collimated outflows emerging from both systems in the form of Herbig-Haro objects \citep{smithetal10, reitersmith13}.
The spatial distribution of the bipolar outflows also corroborates the disk nature of the sources since the disks are located at the origin of each outflow, and their major axes are aligned almost perpendicular to them (Fig.~2).
In the case of 105-600, the presence of an hourglass-shaped near-infrared reflection nebula is additional evidence for a rotationally flattened envelope or disk \citep{sahaietal12}. 

By comparing the density fluxes between the spectral windows at 220~GHz and 237~GHz, spectral indices higher than 2 are determined for the two sources, implying that the observed emission is most likely produced by dust grains (see Table~\ref{tab1}). 
Assuming that the dust is optically thin and isothermal \citep{hildebrand83}, these indices suggest that the grains have grown to larger sizes than are seen in interstellar medium dust \citep{ossenkopfhenning94}, consistent with expectations for grain evolution in the mid-plane of evolved, settled disks \citep{ackeetal04}. 

The total disk mass (\(M\disk\)) can be derived under the same assumptions, being proportional to the mm~flux density \(S_\nu\):
\begin{equation}
  \label{eq:1}
  M\disk = \frac{S_\nu \, D^2}{B_\nu(T_{dust})\, \kappa_\nu},
\end{equation}
where \(D\) is the distance to Carina, \(B_\nu(T_{dust})\) is the Planck function at the dust temperature \(T_{dust}\), and \(\kappa_\nu\) is the absorption coefficient per unit of total (gas \(+\) dust) mass density.
The estimated values for the total masses associated with the dust continuum emission are presented in Table~\ref{tab1}.
We have used a dust mass opacity \(\kappa_{1300\mu m} = 0.015\)~cm\(^2\) g\(^{-1}\) (\citealp{ossenkopfhenning94}, assuming a 100:1 gas-to-dust mass ratio) and \(T_{dust} = 40\)~K, as appropriate for disks around intermediate mass stars \citep[e.g.][]{nattaetal00}. 
For the brighter of the two sources, 104-593, the peak observed flux density of approximately 0.6~mJy/beam is roughly equal to the 230~GHz Planck function at 40~K, implying that the central parts of the disk may be optically thick, leading to the true masses being
underestimated.
The mass estimates are uncertain within a factor of a few due to potential deviations of the dust opacity from our adopted value, depending on the degree of grain growth within the disk, plus uncertainties in the dust temperature (see Table~\ref{tab1}).   

We derive masses of about 50~\(M\Jup\) and 30~\(M\Jup\) for 104-593 and 105-600, respectively. 
{These masses are on the upper end of the typical M\disk\ distribution found in Class I sources in less hostile environments as Taurus, Ophiuchus, and Orion \citep[e.g.][]{andrewswilliams07, eisnercarpenter06, williamsetal13, andrewsetal13, carpenteretal14, mannetal14, cieza15}.
The disks can be considered protoplanetary because the measured masses are well above the minimum \(M\disk\) of about \(10~M\Jup\) required for a pre-solar nebula to develop a planetary system \citep{weidenschilling77}. 
Additionally, the average age of the Carina population (\(\sim 1-4\)~Myr; SB08) perfectly fits the required minimum timescale to form planets (\(\sim 1-2\)~Myr; see \S\ref{intro}).
Planet formation may also occur on shorter scales in Herbig systems \citep{mendigutiaetal12} so it is plausible that young planets are forming or already present within these EGGs.

No millimeter emission was detected above the \(4\sigma\) threshold within either the 104-598 or the Tr~14 fields, yielding an upper limit of \(\sim 7~M\Jup\) to the mass of any disk that might be present (see Table~\ref{tab1}).  
In the case of 104-598, no infrared source is detected \citep{povichetal11, preibischetal11}, implying that the EGG does not (yet) contain a YSO.
Similar starless EGGs are seen in the Cygnus OB2 \citep{wrightetal12} and are especially common in M~16 \citep{hesteretal96, caughreanandersen02}.

{In the Tr~14 region, a total of 12 YSOs are found among candidate proplyds \citep{ascensoetal07, smithetal10} and  sources with near-infrared excess \citep{preibischetal11}. 
Unfortunately, half of them are located near the field edge where the sensitivity is lower (Fig.~\ref{f1}). 
These YSO disks must be much less massive than those detected in 104-593 and 105-600.
If the disks were warmer than 40~K (which may well be the case due to external heating from the nearby massive stars), then the \(4\sigma\) limit would be even lower than \(7~M\Jup\), being roughly similar to the median mass of \(\sim 5-8~M\Jup\) for Class II disks \citep[e.g.][]{andrewswilliams05, eisner12}. 
According to the M\disk\ distribution in low-mass star-forming regions \citep[e.g.][]{williamscieza11}, we expected to detect about half of the YSOs above the threshold.
The high flux of ionizing and dissociating radiation produced by the massive members of Tr~14 might explain why we have not detected massive disks in this field. 

\section{Photo-evaporation of Disks and EGGs} \label{photo}  
Theoretical values of the disk mass photo-evaporation rate \({\dot M}\) can depend on the strength of the illuminating radiation field, the disk size and YSO mass, whether it is non-ionizing FUV or ionizing EUV radiation that controls the evaporation \citep[e.g.][]{johnstoneetal98}.
In the simple case of EUV radiation (\(\lambda < 912~\AA\)), the incident photon flux \(F\EUV\) balances the recombination rate per unit area in the evaporation flow plus the hydrogen ion flux through the ionization front, allowing us to estimate:
\begin{multline}
  \label{eq:3}
  \dot{M}\EUV \simeq 
    4\times 10^{-9} \ 
  \left( \frac{F\EUV}{10^{10} \mathrm{\ s^{-1}\ cm^{2}}}  \right)^{1/2} \
  \left( \frac{R\disk}{100 \mathrm{\ AU} } \right)^{3/2} 
  \ M_\odot \mathrm{\ year^{-1}}.
\end{multline}

A sufficient flux \(F\FUV\) of FUV radiation (from \(912~\AA\) to \(2000~\AA\)) can increase the global \({\dot M}\) via a warm atomic disk wind, which inflates the ionization front to several times \(R\disk\) \citep[e.g.][]{johnstoneetal98}. 
The physics of FUV evaporation is less well understood and can depend on the rate of viscous spreading of the disk \citep{andersonetal13}.  
It is most efficient in the supercritical case \citep[see][]{adamsetal04} when the sound speed in the neutral wind exceeds the gravitational escape speed from the disk, requiring \(T \simeq 1000~K\) or \(F\FUV > 10^4 G_0\) if \(R\disk =
60~\mathrm{AU}\), where \(G_0 = 1.6 \times 10^{-3} \ \mathrm{erg\ s^{-1}\ cm^{-2}}\) is the mean FUV field in the solar neighborhood.  

For the Tr~14 region, where we do not detect any disks, the conditions are certainly extreme. 
The cluster has an UV luminosity \((Q_H)\) of \(2 \times 10^{50} \mathrm{\ photon\ s^{-1}}\) (SB08), implying \(F\EUV= 10^{14} \mathrm{\ photon\ s^{-1}\ cm^{-2}}\) and  \(F\FUV \approx 10^6 G_0\).  
The \({\dot M}\EUV\) alone then gives \(2\times 10^{-7} \ M_\odot~\mathrm{\ year^{-1}}\) for \(R\disk = 60~\mathrm{AU}\), and FUV radiation would likely increase this by factors of 2--4.  
The corresponding evaporation timescale \(M\disk/{\dot M}\) is then as short as \(10^5\)~years for an initial disk mass of \(50~M\Jup\). 
Photo-evaporation becomes inefficient once the disk shrinks to a radius \(R\disk \approx 3 (M_* / M_\odot)~\mathrm{AU}\),  where the escape velocity is of the order of the ionized sound speed.
If \(M\disk \propto R\disk\) (corresponding to a surface density profile \(\sim R^{-1}\)) then the disk mass in this phase will be \(M\disk < 0.003 M_*\).  
Thus, centrally condensed disks may persist in the core of Tr~14 with compact sizes (\(< 10~\)AU) and low masses (\(< 3~M\Jup\)).

Similar timescales are found by simply scaling results from the Orion proplyds, where empirical estimates of \({\dot M}\) have a median value \(\dot{M} = 10^{-7}~M_\odot/\)year \citep{henneyodell99, tsamisetal13}.  
From equation~\eqref{eq:3}, the \(Q_H\) for Tr~14 and Tr~16 would predict \(\dot{M}\EUV > 10^{-7} M_\odot \mathrm{\ year^{-1}}\) for a \(R\disk = 100\)~AU within radii of \(45\arcsec \approx 0.5\)~pc and \(80\arcsec \approx 0.9\)~pc, respectively.  
These EUV-irradiation zones are shown in Fig.~\ref{f1} (solid red circles), although for Tr~16 the picture is more complex than indicated because the high-mass stars are spread over \(10'\) \citep{povichetal11b}.
Adding the FUV contribution, the irradiation zones become larger (dashed-red circles in Fig.~\ref{f1} from the condition \(F\FUV > 10^4 G_0\)).
Any disk remaining in these areas for 1~Myr will have lost approximately \(0.1~M_\odot \approx100~M\Jup\). 
For disks with initial masses and sizes similar to our detected sources, this would be more than sufficient to shrink them below our detection limit.

Analogous physics operates in the evaporation of EGGs. 
The EUV controls \({\dot M}\) in this case, since the FUV-heated layer is thin compared with the EGG radius.  
Assuming a roughly spherical shape, 105-600 has a radius of \(R\EGG \approx 1840~\mathrm{AU}\) and is located at a projected distance of 17~pc from Tr~16, which has a \(Q_H\) of \(6 \times 10^{50} \mathrm{\ photon\ s^{-1}}\) (SB08).
If the true distance is not much greater than the projected one, then the incident flux is \(F\EUV = 6.5 \times 10^{9} \mathrm{\ photon\ s^{-1}\ cm^{-2}}\), implying \(\dot{M} = 2.5 \times 10^{-7} \ M_\odot \mathrm{\ year^{-1}}\).
Within a factor of two, the same result is derived from the H\(\alpha\) surface brightness of the globule's bright rim \citep[measured on the images from the Advanced Camera for Surveys;][]{smithetal10}, which gives a model-independent estimate of \(F\EUV\), albeit subject to uncertainties in the foreground dust extinction.  
The EGG mass is estimated as \(M\EGG = 0.1\)--\(0.3~M_\odot\) \citep{sahaietal12}, leading to a timescale \(M\EGG/\dot{M}\) of approximately 1~Myr.  
For 104-593, the EGG has a similar size, but is found at a projected distance of 3.3~pc from Tr~14, resulting in a larger ionizing flux \(F\EUV \approx 2 \times 10^{11} \mathrm{\ photon\ s^{-1}\ cm^{-2}}\).
No mass estimate is available for this EGG, but if it were similar to 105-600, then it would be evaporated in about \(2 \times 10^5\)~years.

Once the EGGs have been dispersed, then the disks themselves will be subject to external photo-evaporation.  
Taking our observed parameters of \(M\disk = 50~M\Jup \approx 0.05~M_\odot\) and \(R\disk = 60~\mathrm{AU}\), we find \({\dot M}\EUV\) of \(10^{-9}\) to \(10^{-8} \ M_\odot \mathrm{\ year^{-1}}\), corresponding to timescales of 5 to 50 Myr for the two disks.  
The FUV fluxes at the detected disk positions are weak: \(F\FUV \approx 4500\,G_0\) for 104-593 and \(F\FUV \approx 300\,G_0\) for
105-600, resulting in a relatively cool neutral wind (\(T < 300~\mathrm{K}\)) that would have difficulty escaping the stellar gravitational potential.  
Thus, the EUV-derived timescales are accurate if the systems do not move significantly from their current position.

\section{Conclusions} \label{conclu}  
High spatial resolution observations at millimeter wavelengths with ALMA allow us to overcome the problem of stellar crowding in the densely-populated centre of Tr~14.
Therefore, given the number of YSO sources within our observed field (Fig.~1), the non-detection of any disks in the exposed core of the cluster is evidence for accelerated disk dispersal.
This is consistent with the disk fraction, as measured by the near-infrared excess emission among x-ray selected YSOs in Carina, being significantly smaller than in lower mass clusters of a similar age \citep{preibischetal11}. 

Roughly 50\% of the YSO population in Carina is found in sparse small groups that are widely distributed over an area of several square parsecs \citep{preibischetal11b, zeidleretal16}. 
The newly detected Carina disks belong to this distributed population, and are both located at larger distances from the ionizing sources, where external evaporation will not threaten their survival over the next 10~million years. 
Disk formation and evolution in most of these YSOs will probably proceed similarly to those in less massive clusters, and planetary formation should occur without being severely affected by the cluster environment \citep{richertetal15}.  
Only the more clustered YSOs will experience significant external photo-evaporation.
The lack of disks more massive than about \(10~M\Jup\) in this sub-population will potentially affect the formation of planetary systems similar to the Solar system, mostly preventing the building-up of gas giant planets.

\acknowledgments
We thank the constructive suggestions to the anonymous referee. 
This paper makes use of the following ALMA data: ADS/JAO.ALMA\#2015.1.013\-23.S. 
ALMA is a partnership of ESO (representing its member states), NSF (USA) and NINS (Japan), together with NRC (Canada), NSC and ASIAA (Taiwan), and KASI (Republic of Korea), in cooperation with the Republic of Chile. The Joint ALMA Observatory is operated by ESO, AUI/NRAO and NAOJ. 
Our color compositions are based on the programs 12050 and 13391 made with the NASA/ESA HST, and obtained from the HLA, a collaboration between the STScI/NASA, the ST-ECF/ESA and the CADC/NRC/CSA.
A.M.D.~acknowledges support from the FONDECYT project 3140383, and thanks to H.~C\'anovas the stimulating discussions.
L.Z. and W.J.H. acknowledge support from CONACyT and DGAPA-UNAM, Mexico, through grants CB152913 and PAPIIT IN111215. 
T.H.P.~acknowledges support from the FONDECYT Regular grant No.~1121005.

{\it Facilities:} \facility{ALMA}.


\end{document}